# In Silico Trial to test COVID-19 candidate vaccines: a case study with UISS platform

Giulia Russo, Marzio Pennisi, Marco Viceconti and Francesco Pappalardo

*Abstract*—**SARS-CoV-2 is a severe respiratory infection that infects humans. Its outburst entitled it as a pandemic emergence. To get a grip on this, outbreak specific preventive and therapeutic interventions are urgently needed. It must be said that, until now, there are no existing vaccines for coronaviruses. To promptly and rapidly respond to pandemic events, the application of in silico trials can be used for designing and testing medicines against SARS-CoV-2 and speed-up the vaccine discovery pipeline, predicting any therapeutic failure and minimizing undesired effects. Here, we present an in silico platform that showed to be in very good agreement with the latest literature in predicting SARS-CoV-2 dynamics and related immune system host response. Moreover, it has been used to predict the outcome of one of the latest suggested approach to design an effective vaccine, based on monoclonal antibody. UISS is then potentially ready to be used as an in silico trial platform to predict the outcome of vaccination strategy against SARS-CoV-2.**

## I. INTRODUCTION

As the epicenter of Coronavirus disease 2019 (COVID-19) and emerging severe acute respiratory syndrome (SARS) caused by novel Coronavirus (2019-nCoV) spread is making its way across the world, global healthcare finds itself facing tremendous challenges. According to the World Health Organization (WHO) situation report (91st), updated on 20 April 2020, there have been globally 72846 confirmed cases of 2019-nCoV and 5296 cases of death caused by the virus itself [1].

2019-nCoV (also referred to as SARS-CoV-2 or HCoV-19) [2], is the seventh coronavirus known to infect humans along with SARS-CoV, MERS-CoV, HKU1, NL63, OC43 and 229E [3]. While these last four coronaviruses are associated with mild symptoms, SARS-CoV, MERS-CoV and SARS-CoV-2 can cause severe acute respiratory syndrome [4], especially in elderlies, of which men, and those individuals with comorbidities and immunocompromised conditions [5]). Although it is similar to SARS-CoV, SARS-CoV-2 has an improved ability for pathogenicity [6]. In particular, latest evidences during the ongoing pandemic reveal that patients affected by SARS-CoV-2 can progress their clinical picture from fever, cough, ageusia and anosmia, sore throat, breathlessness, fatigue, or malaise to pneumonia, acute respiratory distress syndrome (ARDS) and multi organ dysfunction illness [7]. Significantly, in most critically ill patients, SARS-CoV-2 infection is also associated with a severe clinical inflammatory picture based on a serious cytokine storm that is mainly characterized by elevated plasma concentrations of interleukins 6 (IL-6) [8]. In this scenario, it seems that IL-6 owns an important driving role on the cytokine storm, leading to lung damage and reduced survival [9].

To get a grip on this outbreak and flatten the curve of infection, a specific therapeutic intervention to prevent the severity of the disease is urgently needed to reduce morbidity and mortality because, until now, there are no existing vaccines for coronaviruses.

The ideal profile for a targeted SARS-CoV-2 vaccine must address the need of vaccinating human population, with particular regard of those individuals classified as at high risk, comprising, for example, frontline healthcare workers, individuals over the age of 60 and those that show debilitating chronic diseases.

Recently, specific findings about the genome sequencing of SARS-CoV-2 in different countries where cases of infection were registered, revealed its relative intrinsic genomic variability, its virus dynamics and the related host response mechanisms, unveiling interesting knowledge useful for the formulation of innovative strategies for preventive vaccination.

Specifically, SARS-CoV-2 sequencing along with its relative intrinsic genomic variability [10], the presence of minority variants generated during SARS-CoV-2 replication [11], the involved cellular factors that favors SARS-CoV-2 cell entry [12], the timing in which viral load peaks (during the first week of illness), its gradual decline (over the second week) and the increasing of both IgG and IgM antibodies (around day 10 after symptom onset) represent some of the relevant insights so far delineated and considered by research community about SARS-CoV-2 virus [13].

Even though these findings are having several practice consequences and suggest valuable conclusions, SARS-CoV-2

Submitted on. This work was supported in part by ….". *(Corresponding author: Francesco Pappalardo).*

G. Russo is with the Department of Drug Sciences, University of Catania, Catania, Italy (e-mail: giulia.russo@unict.it).

M. Pennisi is with the Computer Science Institute, DiSIT, University of Eastern Piedmont, Alessandria, Italy (e-mail: marzio.pennisi@uniupo.it).

M. Viceconti is with the Department of Industrial Engineering, Alma Mater Studiorum – University of Bologna, Italy (email: marco.viceconti@unibo.it)

F. Pappalardo is with the Department of Drug Sciences, University of Catania, Catania, Italy (e-mail: francesco.pappalardo@unict.it).

dynamics has not been yet fully understood. Information about which parts of SARS-CoV-2 sequence are recognized by the human immune system is still limited and scarcely available. Such knowledge would be of immediate relevance and great help for the design of new vaccines, facilitating the evaluation of potential immunogenic candidates, as well as monitoring the virus mutation events that would be transmitted through the human population.

Currently, there are at least 42 vaccine candidates around the world under development and evaluation at different stages against COVID-19 [14], also accordingly from what reported by WHO through its continuously undergoing landscapes documents concerning the COVID-19 candidate vaccines. These promising vaccine candidates deal with several vaccine technologies based on recombinant protein subunits [15], nucleic acids [16], non-replicating and replicating viral vectors [17], [18], protein constructs [19], virus-like particles [20], live-attenuated virus strains [21], inactivated virus [14], or human monoclonal antibodies (mAbs) [22].

Today, challenges of continuing development of solutions for COVID-19 pandemic are a mandatory need. As never before, the application of modeling and simulation can actively design better vaccine prototypes, support decision making, decrease experimental costs and time, and eventually improve success rates of the vaccine. To this aim, in silico trials (ISTs) for design and testing medicines [23]–[25] can accelerate and speed-up the vaccine discovery pipeline, predicting a therapeutic failure and minimizing undesired effects.

Beyond traditional modeling techniques or applications, Agent-Based Models (ABMs) represent a paradigm that can cover the entire spectrum of the vaccine development process [26], especially for the quantification and prediction of the humoral and cellular response of a specific candidate vaccine as well as its efficacy [27].

The simulation platform we use from fifteen years, named Universal Immune System Simulator (UISS), is based on agent-based methodology, which is able to brilliantly simulate each single entity of the immune system (and consequently its dynamics), along with the significant immune responses induced by a specific pathogen or stimulus. Recently, UISS provided different success stories in immunology field as it is most widely reported in the literature [28]–[31].

We chose to analyze, within the wide landscape of potential candidate vaccines against SARS-CoV-2, a specific cross-neutralizing antibody that Wang et al. [32] suggest to be promising in targeting and binding a communal conserved epitope of SARS-CoV-2 and SARS-CoV on the spike receptor binding domain [33], through an independent mechanism of receptor binding inhibition.

As a case study, here we report a first application of UISS in silico platform to provide predictions of the efficacy of a potential therapy against COVID-19 based on a mAb strategy intervention like the one proposed by Wang et al.

## II. METHODOLOGY

### A. Introduction to Agent-Based Models and UISS, an in silico platform for the human immune system simulation

Agent-Based Models (ABMs) belong to the class of mechanistic models, a family of models that, differently from data-driven models, uses a description of the underlying mechanisms of a given phenomenon to reproduce it. Such a description is usually based on different observational data, previous knowledge and/or hypotheses, and is usually aggregated and rationalized into a conceptual map (i.e., a flow chart and/or a schematic disease model) that reassumes the cascade of events of the phenomenon under investigation. The conceptual map is then translated into mathematical/computational terms and then executed by computers to observe, in silico, the evolution of the phenomenon over time. Besides ABMs, other modeling techniques based on the mechanistic approach can be used. Among these, we recall, for example, ordinary and partial differential equations [34]–[36] and Petri nets [37], [38].

As the name suggests, agent-based models are based on the paradigm of 'agents', autonomous entities that behave individually according to established rules. Such entities can be heterogeneous in nature, and are usually represented on a simulation space where they are free to move, interact each-other and change their internal state as a consequence of interactions. From a computer science perspective, agents can be seen as stochastic finite-state machines, capable of assuming a limited number of discrete states. Using ABMs, the global evolution of the phenomena is observed by taking into account the sum of the individual behaviors of all agents, and sometimes unexpected "emergent" behaviors may be observed.

ABMs have been successfully applied in many research fields, from social sciences to ecology, from epidemiology to biology. In the field of immunology, we developed the Universal Immune System Simulator (UISS), an agent-based framework that has been extended through the last decades to simulate the behavior of the immune system response when challenged against many diseases.

In UISS agents are used to describe cells and molecules of the immune system, as well as external actors that can destabilize (i.e., pathogens such as viruses and bacteria) or restore (i.e. prophylactic and therapeutic treatments) the normal health of the host.

One of the main features of UISS is its ability to mimic the adaptive immune response mechanisms. Mammals have in fact developed an advanced immune system machinery capable to specifically recognize pathogens in order to better react against them. This advanced response is based on the ability to exactly recognize foreign proteins (i.e., epitopes) on pathogens surface by means of receptors, through a key-to-lock mechanism. While an explicit implementation would be both unfeasible and partially inaccurate from a computational point of view, in UISS we mimic such a process through the use of binary strings. Binary strings are used for both representing epitopes and immune system cells' receptors, and the probability that an immune system cell recognizes a pathogen is proportional to the Hamming distance (the number of mismatching bits) between the two strings involved into the interaction. Although this abstraction may seem binding, millions of interactions can be simulated quickly on modern computers, making easier the reproduction of many features of the immune system such as memory, specificity, tolerance and homeostasis. For example, this abstraction demonstrated able to allow the selection of the best adjuvant among a series of candidates for an influenza vaccine when properly coupled with results coming from existing binding prediction tools [28]. This suggests how such an abstraction is able to capture the complexity of the problem.

Besides of receptors, UISS implements many other immune system mechanisms, as thymus selection, haematopoiesis, cell maturation, Hayflick limit, aging, immunological memory, antibody hyper-mutation, bystander effect, cell anergy, antigen processing and presentation.

Up to now, UISS *in silico* platform has been successfully applied to the design and verification of novel treatments for many diseases in both preclinical and clinical environments, including pathologies such as mammary carcinoma [39] and derived lung metastases [40], melanoma [41], atherosclerosis [42], multiple sclerosis [31] and influenza [28].

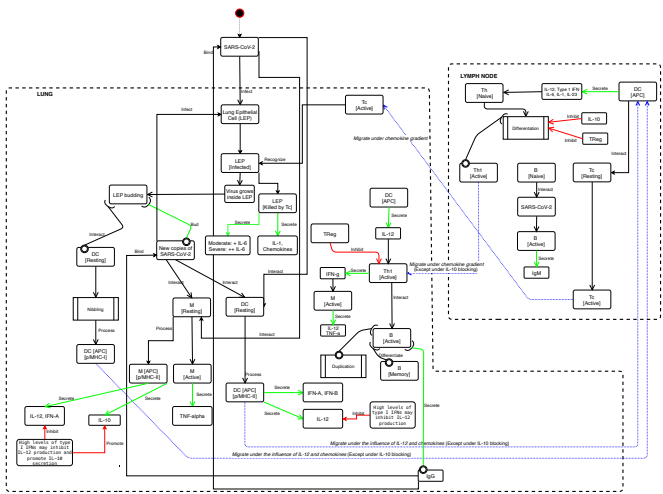

Fig. 1. SARS-CoV-2 disease model implemented in UISS. Main compartments (lung, and lymph-nodes) are delimited with dashed lines. Peripheral blood compartment is seen as connecting duct, not explicitly represented. The starting point is the SARS-CoV-2 droplets entrance in the upper respiratory tract (not shown). Then, all the main infection dynamics is described. The immune system cascade is shown as it was implemented, based on the latest research results published in specialized literature. For each entity, the localization (i.e., the biological compartment in which the entities are present) and the status (i.e., the differentiation states that an entity can own) are defined. The results of the immune system mounting process is the killing of the infected lung epithelial cells by the cytotoxic T lymphocyte and the local release of both chemokine factors and cytokines. At the humoral level, specific IgM (first) and IgG (after) directed against SARS-CoV-2 virus are released by plasma B cells. Regulatory system is also involved in the process. If the immune system machinery works correctly, regulatory arm shutdowns excessive cytokines storm, avoiding the severe prognosis of COVID-19.

More recently, UISS has been used as the centerpiece of the StriTuVaD H2020 project with the aim to create an *in silico* trial for tuberculosis. In this context, observations from virtual patients will be coupled with results from a real clinical trial to obtain an *in silico* augmented clinical trial, with greater accuracy and more statistical power [30].

### B. SARS-CoV-2 disease model

The SARS-CoV-2 disease model has been implemented in UISS computational framework starting by identifying a question of interest. The question of interest describes the specific question, decision or concern that is being addressed with a computational model. In other words, the question of interest lays out the engineering question that is to be answered (at least in part) through a model. The next step is to define the context of use (CoU), which provides a detailed and complete explanation of how the computational model output will be used to answer the question of interest. In this specific study, the question of interest is how potential prophylactic or therapeutic vaccines could cure COVID-19, building or stimulating an effective immune response against SARS-CoV-2 virus. UISS must then represent and reproduce the fundamental SARS-CoV-2 – immune system competition and dynamics. To this end, we first selected all the players that have a role in the viral infection both at cellular and molecular scale and then we categorized all the interactions among entities that play a relevant role in this biological scenario. Finally compartment assumptions have to be done to let the entities move and interact each other. In our case, we considered the lung compartment that models the main organ target of the virus and the generic lymph node that allows immune system entities to be activated and selected. Figure 1 gives a detailed sketch on the main compartments, entities and interactions.

SARS-CoV-2 first entry is located in the upper respiratory tract. Then it proceeds to bronchial and finally to lungs in which it reaches its main cellular target i.e., the epithelial lung cells (LEP) [43]. The virus is eventually captured by dendritic cells (DC) and macrophages (M). DC are the main antigen processing cells of the immune system [44] that are able to present the peptides antigen complexed in both major histocompatibility class I and class II (MHC-I and MHC-II, respectively). If a DC encounters the native virus form, it can be able to process it and present its peptides complexed with MHC-II to CD4 T cells for further actions. DC, upon virus activation, release interferon type A and B (IFN-A and IFN-B) and interleukin-12 (IL-12) that are important cytokines in fighting intracellular pathogens. Also, M are able to capture the native form of the SARS-CoV-2 and, if properly activated by pro-inflammatory cytokines, be able to internally destroy it. After their successful activation, macrophages release a pro-inflammatory cytokine that is tumor necrosis factor alpha (TNF-alpha). A fraction of SARS-CoV-2 viruses reach LEP and through the envelope spike glycoprotein binds to their cellular receptor, ACE2. Doing that, the viral RNA genome starts to be released into the cytoplasm and is translated into two polyproteins and structural proteins, after which the viral genome begins to replicate inside the cell [45]. Following the flux of the conceptual disease model represented in Figure 1, after a certain amount of time (that we tuned with available data, as described in the next sections), new copies of the virus are released from the infected LEP that eventually dies. New released copies of functional SARS-CoV-2 infect new cells, spreading further the infection in the lungs. When a cell is infected by a virus, it can be susceptible of different destinies. One of them is the shutting down of MHC-I expression to avoid immune system recognition from specific CD8 T cells. In this case, a population of innate immunity cells, natural killer cells (NK) may identify them and proceed to kill them through specific actions. The other one is represented by a different MHC-I presentation on the cell surface, as the virus has modified the normal behavior of cell to let the host to make functioning virus copies. In this circumstance, (that we supposed to happen during SARS-CoV-2 infection) cell MHC-I presentation is different from the normal case. DC are able (through a mechanism known as "nibbling" process [46]) to cross present the antigen complexed with MHC-I proteins to let adaptive immune response to recognize and kill virus infected cells. Activated and antigen presenting cells (both DC and/or M) migrate into the proximal lymph nodes to present their content to adaptive immune cells i.e., T cells and B cells. Also, a portion of viruses could eventually migrate to the lymph nodes. Here, B cells can be activated by virus if specific immunoglobulin receptor in B cell surface binds to it. In this context, B cell is activated, and it immediately releases immunoglobulins of M class (IgM) that are the first antibody response that can be measured. Further, APC cells activate CD4 T cells (helper T cells, Th) that under the influence of specific cytokines released before, differentiate into helper T cell type 1 (Th1). Th1 migrate under chemokines gradient to the site of infection. There, they release interferon gamma (IFN-G) that makes macrophages able to destroy captured viral particles and allow them to release IL-12 that promotes immune system activation against the virus. Th1 cells allow the differentiation and the iso-switching B cells into immunoglobulins class G (IgG) producing plasma cells. IgG are specific antibodies that bind against virus receptors, eventually inhibiting its capacity to infect cells. MHC-I/peptides DC presenting cells are also able to activate CD8 cytotoxic T cells (Tc) to destroy SARS-CoV-2 infected cells and then eliminate the reservoir of infection. Eventually, Tc migrate into the site of infection and recognize and kill infected LEP. Tc killed infected LEP release chemokines and interleukin 1 and 6 (IL-1 and IL-6). IL-1 is the main cytokine that induces several systemic effects in the host, for example fever. IL-6 is a proinflammatory cytokine that can change the severity of COVID-19 disease as reported in very recent literature [47]. Our disease model takes good account of the cytokines storm in the prognosis of the severity of the disease. Entities (both cellular and molecular) move and diffuse in a simulation space represented as a L X L lattice (L is set depending on the dimension of the compartment one intends to reproduce), with periodic boundary conditions. There is no correlation between entities residing on different sites at a fixed time as the interactions among cells and molecules take place within a lattice-site in a single time step. All entities are allowed to move with a uniform probability between neighboring lattices in the grid and with an equal diffusion coefficient (Brownian motion).

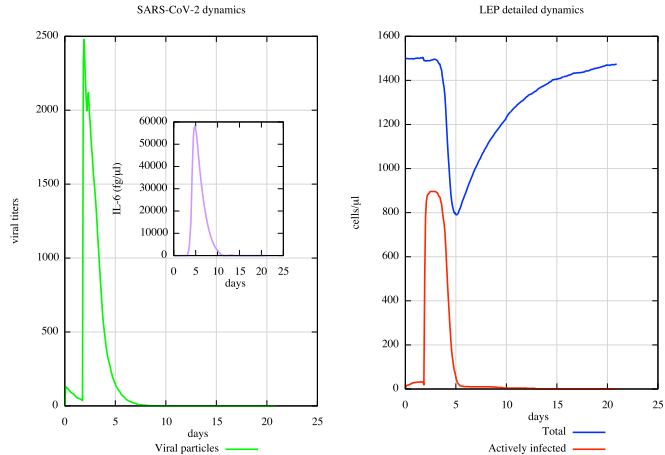

Fig. 2. In silico SARS-CoV-2 viral dynamics and related CPE in a mild to moderate scenario. In the left panel, one can observe the mild digital patient case in which a virus challenge dose of 0.1 multiplicity of infection (MOI) was administered at day 0 (green line). Peak viral titers are reached by 48 h post-inoculation. IL-6 dynamics and its related plasma levels (fg/μL) are also shown in the inner panel (purple line). In the right one, the dynamics of CPE on the lung infected cells is measured: they started at day 3.5 and peak around day 5. After 21 days, the simulated digital patient almost recovers from the infection. One can notice how UISS is capable to simulate, accordingly to the recent literature, the early viral clearance by day 10 post-onset in mild cases.

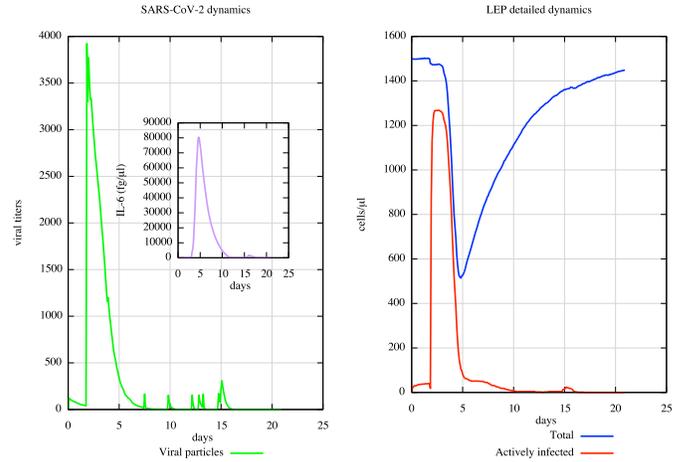

Fig. 3. In silico SARS-CoV-2 viral dynamics and related CPE in a severe scenario. In the left panel, one can observe the severe digital patient case in which a virus challenge dose of 0.1 multiplicity of infection (MOI) was administered at day 0 (green line). Peak viral titers are reached by 48 h post-inoculation. In addition, it is wort to note that virus persists after day 10, until day 15, and its complete clearance is around day 19. In the inner panel (purple line), IL-6 dynamics and its related plasma levels (fg/μL) are shown. IL-6 dynamics shows a much more prominent peak of values. This is in very good agreement with latest literature data, as explained within the manuscript. In the right panel, the dynamics of CPE on the lung infected cells is measured: in this case, CPE are much more severe and the recover from infection is clearly delayed. UISS is capable to simulate, accordingly to the recent literature, how the severe cases tend to have a higher viral load both at the beginning and later on.

## III. RESULTS AND DISCUSSION

### A. Tuning and validation of SARS-CoV-2 disease model

Scientific knowledge about SARS-CoV-2 is still not complete and research contributions appear every day. Apart from this, we used all the available literature data to compare the dynamics predicted by the UISS platform with all findings we were able to fetch. The first task we accomplished with success was the evaluation of the replication kinetic of SARS-CoV-2. To this end, we set a first use case simulation considering a digital patient in which a virus challenge dose of 0.1 multiplicity of infection (MOI) was administered at day 0.

Simulation space was 5 cubic millimeters of lung tissue, 5 cubic millimeters of lymph tissue and 5 microliters of peripheral blood. Figure 2 (right panel) shows that peak viral titers are reached by 48 h post-inoculation. We also plotted IL-6 dynamics: as reported in [9] the levels of IL-6 could be provide a prognosis on the severity of the infection.

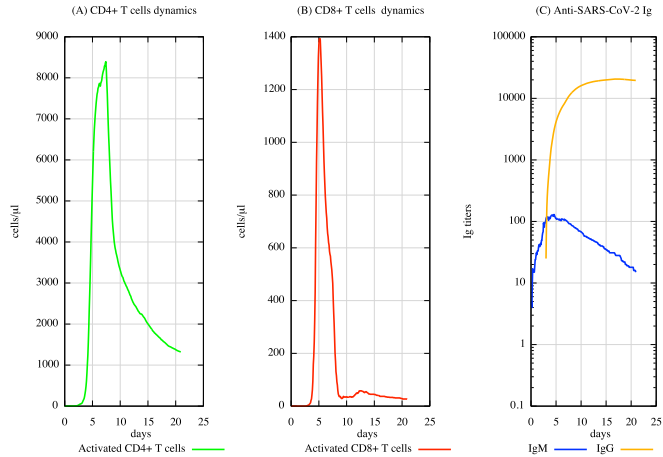

Fig. 4. Cellular and humoral response mounted by the host immune system against SARS-CoV-2. Panel A shows the dynamics of CD4$^+$T cells, subtype 1 (Th1). Th1 are primed by dendritic cells that present the viral particles complexed with MHC-II of the host. Th1 cells help the activation of B cells, eventually favoring their iso-type switching to IgG producing plasma cell. B cells dynamics is depicted in panel B. Antigen activated B cells initially releases IgM; then, after interacting with Th1 and their released pro-inflammatory cytokines, they start to release specific IgG directed against SARS-CoV-2 virus.

We also measured cytophatic effects (CPE) on the lung compartment. CPE are defined as changes occurred in the infected cell that eventually lead its lysis or inability to reproduce. Figure 2 (left panel) highlights the dynamics of CPE: they started at day 3.5 and peak around day 5. After 21 days, the simulated digital patient almost recovers from the infections. These findings are in good agreement with actual literature [13], [48].

In a recent work, Liu et al [49] reported that mild cases were found to have an early viral clearance, with 90% of these patients repeatedly testing negative by day 10 post-onset. At the same time, they found that all severe cases still tested positive at or beyond day 10 post-onset. Moreover, severe cases tended to have a higher viral load both at the beginning and later. In contrast, mild cases had early viral clearance 10 days post onset. UISS was also able to reproduce this scenario. As one can see, Figure 2 is in very good agreement for viral clearance.

We also were able to reproduce severe conditions acting on the immune system aging parameters obtaining results showed in Figure 3. In this case, Figure 3 (right panel) shows virus presence after day 10, until day 15, and its complete clearance about day 19. Moreover, CPE are much more severe and the recover from infection was clearly delayed (left panel). IL-6 dynamics shows a much more prominent peak of values. This is in very good agreement with latest literature data, as explained before.

To validate the main immune system response of mild-to-moderate COVID-19 patients, we used the results available in [50]. In this work, the authors report the kinetics of immune responses in terms of activated CD4$^+$ T cells, CD8$^+$ T cells, IgM and IgG antibodies, detected in blood before symptomatic recovery. As one can see from Figure 4, the kinetics of activated Th1 cells (panel A), activated CD8 T cells (panel B) and the IgM and IgG (panel C) predicted by the simulator are in good agreement with their findings.

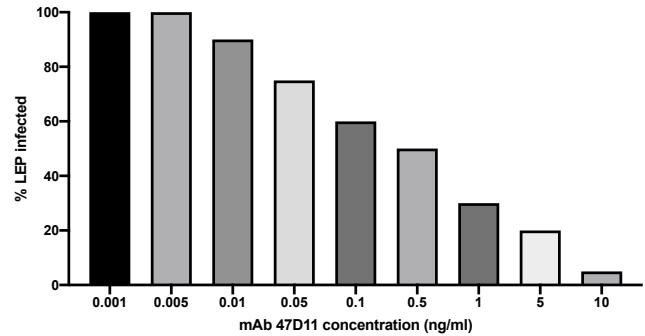

Fig. 5. Antibody-mediated neutralization of SARS-CoV-2 infection on simulated lung epithelial cells. 10ng/ml revealed the best concentration to obtain maximum clearance of the virus.

### B. UISS IST to predict mAb efficacy against SARS-CoV-2

UISS is an immune system simulation platform that was designed to be applied to several and different scenarios, especially to carry on in silico trials to predict the efficacy of a specific prophylactic or therapeutic vaccine against a particular disease. In silico trials aim to strongly reduce the time to develop new effective therapeutics: this is particularly crucial in situation like the one we are facing with. As soon as a disease model incorporated into UISS is tuned and validated against available data, it can be used as an *in silico* lab to test new vaccines. In the previous section, we demonstrated that UISS-SARS-CoV-2 is able to reproduce and predict the main aspects of the viral infection. As a working example, here we show how the platform can be immediately used to predict the efficacy of a human monoclonal antibody that neutralizes SARS-CoV-2 developed by Wang et al. [32]. In this work the authors suppose that the developed antibody (named 47D11) neutralizes SARS-CoV-2 through a yet unknown mechanism that is different from receptor binding interference. Hence, in implementing the mechanism of action of 47D11 into UISS computational framework we used the alternative mechanisms of coronavirus neutralization by receptor-binding domain (RBD) targeting antibodies that have been reported, including spike inactivation through antibody-induced destabilization of its prefusion structure, which Wang et al. indicated also applicable for 47D11. We then modeled the receptor interaction to trigger irreversible conformational changes in coronavirus spike proteins enabling membrane fusion, as described in [51]. The validation in silico trial consists in simulating the in vitro experiment conducted by Wang et al. where they showed that the monoclonal antibody was effective in contrasting SARS-CoV-2 to infect the target cells. To mimic the in vitro system that is an isolated system, we disabled both the lung and the lymph node compartments and we turned off all the immune system interactions in UISS. We then injected virus particles in the peripheral blood compartment of the simulator along with different concentrations of mAb. Only LEP-SARS-CoV-2 interaction has been allowed to happen. Figure 5 shows the obtained results.

As one can envisage from the figure, we reproduced with great accuracy the in vitro results reported by Wang et al. In particular, the computational framework was able to correctly predict the efficacy of 47D11 mAb simulating its mechanism of action that induces the entering of the virus inside LEP. The most effective concentration is 10 ng/ml. This step makes the simulation platform ready for usage as a in silico trial to predict now the effects of a mAb therapy. For this purpose, we designed two kind of in silico experiments. The first one dealt with mAb vaccination strategy used to prevent the onset of infection. The second one involved mAb as interventional drug to treat already infected hosts.

### 1) mAb as a preventive vaccine

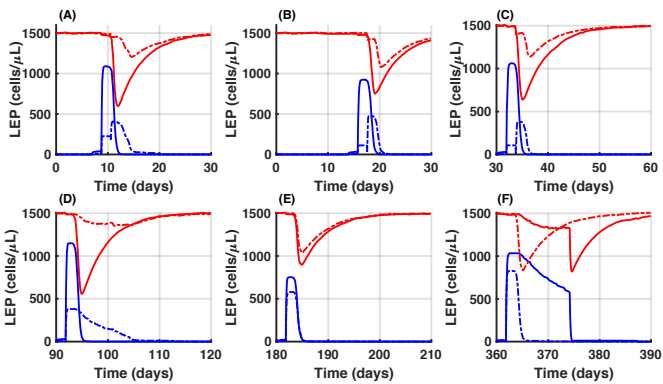

Fig. 6. In silico trial of 47D11 to predict preventive efficacy. The overall prediction dynamics of LEP at the injection time of 47D11 mAb (day 1) at a concentration of 10ng/ml after different SARS-CoV-2 challenge in time (panels A to F) is depicted. Specifically, one can observe the exposures at virus particles at day 7 (panel A), day 14 (panel B), month 1 (panel C), month 3 (panel D), month 6 (panel E) and after 1 year (panel F). Solid lines refer to the no-treated digital patient while dashed lines refer to mAb treated one. Blue lines depict actively infected LEP while red lines represent LEP to show the CPE. Panel A highlights that if a patient enters in contact with the virus just after 7 days post vaccination, he is fully protected from the infection (SARS-CoV-2 actively infected values of LEP are low). This is also true if a potential subject is being infected after two weeks (panel B), one month (panel C) or three months (panel D) after vaccination. Oppositely, in panel E (subject infected after 6 months) and panel F (subject infected after one year) mAb vaccination is practically ineffective in protecting the onset of the disease.

Figure 6 depicts the dynamics of LEP while we injected the 47D11 mAb at day 1 at a concentration of 10ng/ml. Subsequently, we injected SARS-CoV-2 virus particles at day 7 (panel A), day 14 (panel B), month 1 (panel C), month 3 (panel D), month 6 (panel E) and after 1 year (panel F). Solid lines refer to the no-treated digital patient while dashed lines refer to mAb treated one. Blue lines depict actively infected LEP while red lines represent LEP to show the CPE.

Panel A of figure 6 clearly demonstrates that if a patient enters in contact with the virus just after 7 days post vaccination, he is fully protected from the infection: as one can see, SARS-CoV-2 actively infected values of LEP are low. This is also true if a potential subject is being infected after two weeks (panel B), one month (panel C) or three months (panel D) after vaccination. Things change for the other two cases i.e., panel E (subject infected after 6 months) or panel F (subject infected after one year). In this circumstance, the computational framework predicts that the mAb vaccination is practically ineffective in protecting the onset of the disease. A second injection of the mAb is suggested around month 4 to extend the protection of the host for one year.

### 2) mAb as a therapeutic vaccine

The second experiment we designed is to use our in silico trial platform to predict the efficacy of a mAb-based vaccine in therapeutic settings. We envisaged both mild-moderate case (the same digital patient type shown in Figure 2) and severe case (the same digital patient shown in Figure 3). For both cases, we administered the mAb vaccine one day and two days after the onset of infection. Figure 7 shows different behavior for both mild-moderate case (panel A, mAb injected one day after the onset of infection, and panel B, mAb injected two days after the onset of infection) and severe case (panels C, mAb injected one day after the onset of infection, and panel D, mAb injected two days after the onset of infection). As Figure 7 depicts, mAb vaccine is effective in preventing or strongly limiting the CPE. If the mAb vaccine is injected after two days, it is not able to protect the LPE of the host to be infected by the virus and consequently from COVID-19 pathology.

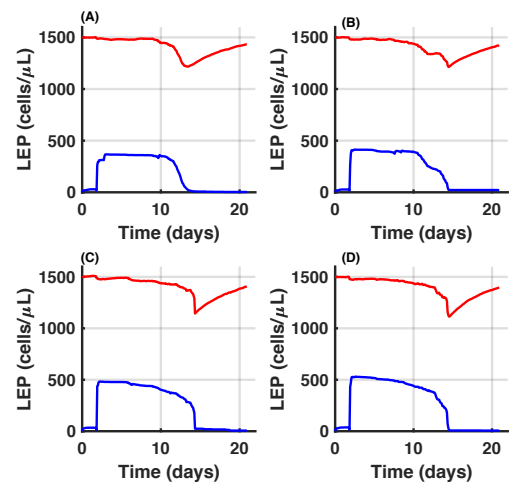

Fig. 7. In silico trial of 47D11 to predict therapeutic efficacy. Different behaviors for both mild-moderate case (panels A and B) and severe case (panels C and D) are shown. In panel A, mAb is injected one day after the onset of infection. In panel B, mAb is injected two days after the onset of infection. In panel C, mAb is injected one day after the onset of infection. In panel D, mAb is injected two days after the onset of infection. Blue lines depict actively infected LEP while red lines represent LEP to show the CPE. As one can notice, mAb vaccine is effective in preventing or strongly limiting the CPE only by two days of the infection.

## IV. CONCLUSIONS

In this paper, we present an *in silico* platform that was demonstrated able to reproduce the main dynamics of SARS-CoV-2 virus and the elicited host immune response against it. The disease model was implemented inside UISS computational framework, an *in silico* trial platform that has been applied to several biological scenario. UISS shows that the simulated SARS-CoV-2 dynamics is in very good agreement with the one described in the latest literature; also, the immune system response predicted by UISS against the virus mirrored the one observed in state of the art research data. This validation step entitled UISS-SARS-CoV-2 to be used as a in silico lab to test the efficacy of potential vaccines for COVID-19, knowing a priori their mechanism of action. Hence, we set an in silico trial to test a recent vaccination strategy based on the employment of monoclonal antibodies directed against a specific target protein of the virus. The simulator is in good agreement in predicting the in vitro experiment outcome performed by the inventors of 47D11 mAb. Finally, we designed two experimental settings to predict the efficacy of mAb vaccination when used in both preventive and therapeutic cases. We predicted that mAb is an effective therapy when used as a preventive vaccine (granting up one year protection when injected with a two times schedule). Moreover, we envisaged that 47D11 mAb if effective only if administered in a very stringent time-frame if employed as a therapeutic strategy.

## ACKNOWLEDGEMENT

The authors would like to thank all the researchers that are involved in finding potentially effective vaccines candidates against SARS-CoV-2 virus. We will be more than happy to collaborate with any of them to support the process.